\documentclass[sigconf]{acmart}

\usepackage{xcolor} 
\usepackage{enumitem} 
\usepackage{amsthm} 
\usepackage{bm} 
\usepackage[ruled,linesnumbered,vlined]{algorithm2e} 
\usepackage{listings} 
\usepackage{array} 
\usepackage{multirow} 
\usepackage{makecell}

\usepackage[authormarkup=none]{changes} 

\graphicspath{{figures/}}
\newcommand{\figref}[1]{Figure~\ref{#1}}
\newcommand{\tabref}[1]{Table~\ref{#1}}
\newcommand{\secref}[1]{Section~\ref{#1}}
\newcommand{\algref}[1]{Algorithm~\ref{#1}}
\newcommand{\listref}[1]{Listing~\ref{#1}}

\newcounter{evaluation}

\newcommand{\evaluation}[2]{\refstepcounter{evaluation} \label{evaluation:#1}
  \begin{center}
  \begin{tikzpicture}%
    \node[rectangle, draw=black, top color=black!3, bottom
    color=black!3, rounded corners=2pt, inner xsep=5pt, inner
    ysep=6pt, outer ysep=10pt]{
    \begin{minipage}{0.95\columnwidth}
      \textit{#2}
    \end{minipage}};%
  \end{tikzpicture}%
  \end{center}
}

\newcounter{observation}

\newcommand{\observation}[2]{\refstepcounter{observation} 
  \label{observation:#1}
  \begin{center}
  \begin{tikzpicture}%
    \node[rectangle, draw=black, top color=black!3, bottom
    color=black!3, rounded corners=2pt, inner xsep=5pt, inner
    ysep=3pt, outer ysep=4pt]{
    \begin{minipage}{0.95\columnwidth}
      \textit{\textbf{O\arabic{observation}}: #2}
    \end{minipage}};%
  \end{tikzpicture}%
  \end{center}
}

\definecolor{darkgreen}{RGB}{0,168,0}
\definecolor{darkpurple}{RGB}{200, 30, 225}

\newcommand{\approach}{\texttt{RISE}}

\definechangesauthor[color=darkpurple]{xudong}

\definechangesauthor[color=red]{wensheng}

\definechangesauthor[color=purple]{yuwei}

\definechangesauthor[color=cyan]{jiansen}

\definechangesauthor[color=orange]{yangrui}

\definechangesauthor[color=brown]{ziyu}

\definechangesauthor[color=pink]{yu}

\copyrightyear{2026}
\acmYear{2026}
\setcopyright{cc}
\setcctype{by}
\acmConference[ICSE '26]{2026 IEEE/ACM 48th International Conference on Software Engineering}{April 12--18, 2026}{Rio de Janeiro, Brazil}
\acmBooktitle{2026 IEEE/ACM 48th International Conference on Software Engineering (ICSE '26), April 12--18, 2026, Rio de Janeiro, Brazil}
\acmPrice{}
\acmDOI{10.1145/3744916.3773257}
\acmISBN{979-8-4007-2025-3/26/04}

\begin{document}

\title{RISE: Rule-Driven SQL Dialect Translation via Query Reduction}

\author{Xudong Xie}
\orcid{0009-0003-3367-5995}
\authornote{Affiliated with Key Lab of System Software at
CAS, State Key Lab of Computer Science at Institute of Software at CAS, and
University of CAS, Beijing. CAS is the abbreviation of Chinese Academy of
Sciences.}
\affiliation{
  \institution{Institute of Software at CAS, China}
  \city{}
  \country{}
}
\email{xiexudong23@otcaix.iscas.ac.cn}

\author{Yuwei Zhang}
\orcid{0009-0008-1016-7361}
\authornotemark[1]
\authornotemark[3]
\affiliation{
  \institution{Institute of Software at CAS, China}
  \city{}
  \country{}
}
\email{zhangyuwei@iscas.ac.cn}

\author{Wensheng Dou}
\orcid{0000-0002-3323-0449}
\authornotemark[1]
\authornote{Affiliated with Nanjing Institute of Software Technology, University of Chinese Academy of Sciences, Nanjing.}
\authornote{Yuwei Zhang and Wensheng Dou are the corresponding authors.}
\affiliation{
  \institution{Institute of Software at CAS, China}
  \city{}
  \country{}
}
\email{wensheng@iscas.ac.cn}

\author{Yu Gao}
\orcid{0009-0009-8312-6736}
\authornotemark[1]
\affiliation{
  \institution{Institute of Software at CAS, China}
  \city{}
  \country{}
}
\email{gaoyu15@otcaix.iscas.ac.cn}

\author{Ziyu Cui}
\orcid{0009-0004-7462-194X}
\authornotemark[1]
\affiliation{
  \institution{Institute of Software at CAS, China}
  \city{}
  \country{}
}
\email{cuiziyu20@otcaix.iscas.ac.cn}

\author{Jiansen Song}
\orcid{0009-0003-0401-2033}
\authornotemark[1]
\affiliation{
  \institution{Institute of Software at CAS, China}
  \city{}
  \country{}
}
\email{songjiansen20@otcaix.iscas.ac.cn}

\author{Rui Yang}
\orcid{0009-0009-5428-2852}
\authornotemark[1]
\affiliation{
  \institution{Institute of Software at CAS, China}
  \city{}
  \country{}
}
\email{yangrui22@otcaix.iscas.ac.cn}

\author{Jun Wei}
\orcid{0000-0002-8561-2481}
\authornotemark[1]
\authornotemark[2]
\affiliation{
  \institution{Institute of Software at CAS, China}
  \city{}
  \country{}
}
\email{wj@otcaix.iscas.ac.cn}

\begin{abstract}

Translating SQL dialects across different relational database
management systems (RDBMSs) is crucial for migrating RDBMS-based applications to the cloud.
Traditional SQL dialect translation tools rely on 
manually-crafted rules, necessitating significant manual effort to support new RDBMSs and dialects. Although large language models (LLMs) can assist
in translating SQL dialects, they often struggle with lengthy and complex SQL
queries.

In this paper, we propose \approach{}, a novel LLM-based SQL dialect translation
approach that can accurately handle lengthy and complex SQL queries. Given a
complex source query $Q_c$ that contains a SQL dialect $d$, we first employ a
dialect-aware query reduction technique to derive a simplified query $Q_{s}$
by removing $d$-irrelevant SQL elements from $Q_c$.
Subsequently, we utilize LLMs to translate $Q_{s}$ into
$Q_{s^{'}}$, and automatically extract the translation rule
$r_d$ for dialect $d$ based on the relationship between $Q_{s}$ and
$Q_{s^{'}}$. By applying $r_d$ to $Q_c$, we can effectively
translate the dialect $d$ within $Q_c$, thereby bypassing the complexity of the
source query $Q_c$. We evaluate \approach{} on two real-world benchmarks, i.e., TPC-DS and SQLProcBench, comparing its performance against both the traditional rule-based tools and the LLM-based approaches with respect to translation accuracy. \approach{} achieves accuracies of 97.98\% on TPC-DS and 100\% on SQLProcBench, outperforming the baselines by an average improvement of 24.62\% and 238.41\%, respectively.

\end{abstract}



\maketitle

\section{INTRODUCTION}

Nowadays, more than 50 relational database management systems (RDBMSs) are widely used across organizations \cite{DBEngines}, many of which have been deployed on the cloud to enhance performance and reduce costs. Different RDBMSs typically offer distinct advantages; for example,
some excel in scalability, while others excel in security. To achieve these benefits, legacy RDBMS-based applications are
expected to migrate
into suitable cloud RDBMSs \cite{CSMM2024, Mallet2024, Zmigrod}.

Although many RDBMSs (e.g., Oracle \cite{oracle2023doc}, MySQL \cite{mysql2023doc},
PostgreSQL \cite{postgres-manual} and SQLite \cite{SQLinesMigration}) support
Structured Query Language (SQL), they incorporate vendor-specific extensions and
unique nuances, resulting in a wide variety of SQL dialects. The presence of
these diverse SQL dialects introduces both syntax and semantic discrepancies, making
SQL dialect translation across different RDBMSs a challenging problem. Effective
SQL dialect translation should ensure semantic equivalence before and after
translation, meaning that the same query results and database states should be returned. Existing study \cite{aws2022} has
demonstrated that SQL dialect translation typically consumes 20-40\% of
migration budgets.

\begin{figure*}[t]
  \centering
  \includegraphics{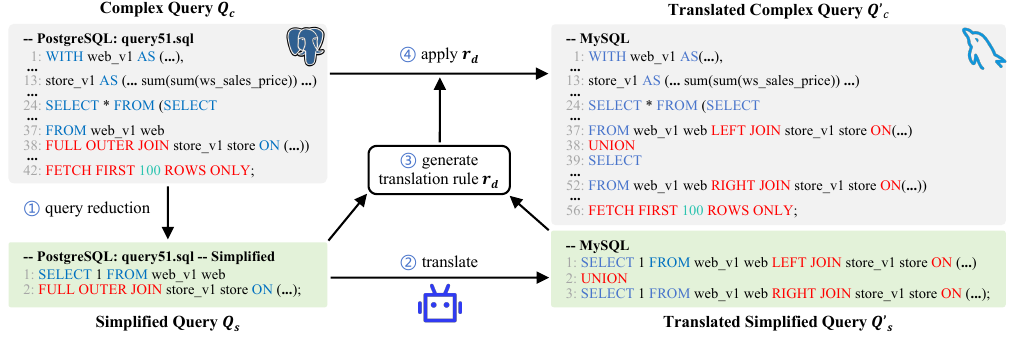}
  \caption{An Example of SQL Dialect Translation.
  PostgreSQL-Specific Dialects are Colored in {\color{red}Red}.}
  \label{fig:motivating-example}
\end{figure*}

Several rule-based tools \cite{Datometry2023, JOOQSQLTranslation, SQLGlot,
SQLinesMigration} have been developed to support SQL dialect translation across
RDBMSs. The widespread adoption of tools (e.g., SQLGlot \cite{SQLGlot} with
7.8k GitHub stars) and proprietary solutions from Microsoft and Amazon
\cite{MicrosoftAzureDSCT, AmazonAWSSchemaTool} highlight the critical need for
addressing SQL dialect translation challenges in both academia and industry
\cite{Zmigrod}. These tools mainly rely on manually-crafted
translation rules, which are often incomplete and error-prone. For example, the
complex query $Q_c$ shown in \figref{fig:motivating-example} cannot be correctly
translated by existing tools. Specifically, jOOQ and SQLines do not support
the \texttt{FULL} \texttt{OUTER} \texttt{JOIN} clause, while SQLGlot generates
an incorrect query by substituting \texttt{UNION} \texttt{ALL} for
\texttt{UNION}, resulting in inconsistent query outcomes. Moreover, the manual
effort required to develop and maintain these translation rules is considerable.
For example, SQLGlot includes 1900 lines of code dedicated to
MySQL-to-PostgreSQL translation.

Recently, researchers have explored the use of large language models (LLMs) to
reduce dependence on manually-crafted rules in traditional tools
\cite{Datometry2023, JOOQSQLTranslation, SQLGlot, SQLinesMigration}. For
example, Mallet \cite{Mallet2024} employs LLMs to generate translation rules,
which are then manually integrated into SQLGlot \cite{SQLGlot} by developers.
Therefore, Mallet inherits many of the limitations associated with SQLGlot.
A more direct approach involves prompting LLMs to translate SQL queries,
referred to as \texttt{LLMTranslator}, which is an experimental prototype
designed in this paper. However, \texttt{LLMTranslator} struggles
with lengthy and complex queries due to hallucination issues commonly observed
in LLMs \cite{Farquhar2024}. For example, when translating the complex query
$Q_c$ from \figref{fig:motivating-example}, \texttt{LLMTranslator} incorrectly
rewrites `\texttt{sum(sum(ws$\_$sales$\_$price))}' as
`\texttt{sum(ws$\_$sales$\_$price)}', altering the intended semantics of the
query.

To assess the practical limitations of \texttt{LLMTranslator}, we conduct an empirical study using two benchmarks (i.e., TPC-DS \cite{tpc_ds_3_2_0} and SQLProcBench \cite{plsql}), both of which are derived from real-world workloads. 
The results demonstrate that \texttt{LLMTranslator} performs well on simple queries, accurately translating SQL dialects with straightforward syntax and semantics. However, its effectiveness declines markedly when handling lengthy and complex queries. In such cases, \texttt{LLMTranslator} frequently exhibits hallucination issues, leading to incorrect translations.
These findings suggest that while \texttt{LLMTranslator} excels in simple translation tasks, it struggles with complex queries due to the presence of dialect-irrelevant noise that can obscure the essential semantics.
Therefore, isolating and filtering out such non-essential elements is crucial for improving the accuracy of LLM-based SQL dialect translation.

Inspired by the above observations, we propose \approach{}, an automated
\uline{\textbf{R}}ule-driven SQL d\uline{\textbf{I}}alect
tran\uline{\textbf{S}}lation approach that utilizes dialect-aware query
r\uline{\textbf{E}}duction. The core idea is to simplify complex SQL queries by
removing dialect-irrelevant elements. This reduction technique produces
simplified queries that retain only the elements specific to the target SQL
dialect. As illustrated in \figref{fig:motivating-example}, the source query
$Q_c$ spans 42 lines, features complex structures, and includes two
PostgreSQL-specific dialects (colored in red). Through our reduction technique,
$Q_{c}$ is first transformed into a simplified query $Q_{s}$, which isolates the
dialect-specific elements of {\texttt{FULL} \texttt{OUTER} \texttt{JOIN}} in two
lines. We then prompt LLMs to translate $Q_{s}$ into $Q_{s^{'}}$, a task that
LLMs can handle effectively due to the reduced complexity. From the
correspondence between $Q_{s}$ and $Q_{s^{'}}$, we generate a translation rule
$r_d$ for the first dialect {\texttt{FULL} \texttt{OUTER} \texttt{JOIN}}.
Finally, we apply $r_d$ back to the original $Q_c$, enabling accurate and
efficient translation of dialect-specific elements without requiring LLMs to
process the entire complex query.

We evaluate \approach{} on two real-world benchmarks, i.e., TPC-DS and
SQLProcBench, which require SQL dialect translations from PostgreSQL to MySQL
and Oracle, respectively. We compare \approach{} against state-of-the-art
baselines, including traditional rule-based tools (i.e., SQLines, jOOQ and
SQLGlot) and LLM-based methods (i.e., our prototype \texttt{LLMTranslator}
and CrackSQL). Experimental results show that \approach{}
substantially surpasses all baselines in terms of translation accuracy. We have
made our tool available at
\href{https://figshare.com/s/5c709f5c0a63d58b0ef4}{https://figshare.com/s/5c709f5c0a63d58b0ef4}.

In summary, we make the following contributions.

\begin{itemize}[noitemsep,topsep=0pt,parsep=0pt,partopsep=0pt,leftmargin=15pt]
  \item We conduct the first empirical study to assess the effectiveness and
  limitations of \texttt{LLMTranslator}, offering valuable insights for
  improving LLM-based SQL dialect translation.
  \item We propose \approach{}, a novel LLM-based SQL dialect approach capable
  of accurately handling lengthy and complex SQL queries. \approach{}
  integrates an effective dialect-aware query reduction technique with an
  LLM-assisted translation rule generator to overcome the limitations of naive
  LLM translation.
  \item We perform extensive experiments on two real-world benchmarks and
  thoroughly evaluate each component of \approach{}. The experimental results
  demonstrate that \approach{} outperforms state-of-the-art baselines,
  highlighting its superior performance.
\end{itemize}

\section{EMPIRICAL STUDY}
Given the extensive knowledge and strong natural language understanding
capabilities of LLMs, leveraging them for SQL dialect translation offers a
highly promising and flexible approach. To assess the effectiveness of LLMs in
translating SQL dialects, we implement a prototype \texttt{LLMTranslator} and
conduct an empirical study aimed at addressing the following two research
questions.
\begin{itemize}[noitemsep,topsep=0pt,parsep=0pt,partopsep=0pt,leftmargin=15pt]
  \item \textbf{RQ1 (Translation accuracy):} Can \texttt{LLMTranslator}
  accurately translate SQL dialects?
  \item \textbf{RQ2 (Failure factors):} What factors can lead to failures in SQL
  dialect translation when using \texttt{LLMTranslator}?
\end{itemize}

\subsection{Study Methodology} 
\label{sec:empirical-study}

We introduce the benchmarks, methodology, and evaluation metric used in the
empirical study below.

\subsubsection{Benchmarks} \label{sec:benchmarks}
We evaluate the capability of LLMs in SQL dialect
translation using queries from the following two benchmarks:
\begin{itemize}[noitemsep,topsep=0pt,parsep=0pt,partopsep=0pt,leftmargin=15pt]
  \item \textbf{TPC-DS:} TPC-DS \cite{tpc_ds_3_2_0} is an industry-standard
benchmark for evaluating the performance of RDBMSs on complex analytical
queries. It consists of 99 queries covering a wide range of SQL constructs,
with some queries exceeding 100 lines. These queries include basic operators,
aggregations, functions, and advanced SQL features (e.g., \texttt{WITH} clauses,
\texttt{CTE}, sub-queries). TPC-DS captures the intricacies and
diversity of real-world query workloads, with each query meticulously designed
to simulate practical analytical tasks. Thus, we employ TPC-DS to evaluate the
ability of LLMs in addressing the multifaceted challenges of industrial-scale
SQL dialect translation.
  \item \textbf{SQLProcBench:} SQLProcBench \cite{plsql} extends the schema
  of TPC-DS by emphasizing procedural code, including stored procedures,
  functions, and triggers implemented in PostgreSQL, Oracle, and SQL Server. In
  our evaluation, we focus on stored procedures, which comprise 44 statements
  designed to simulate real-world procedural code usage in various applications,
  with some exceeding 50 lines of SQL code. These procedures
  exhibit complex control flows (e.g., \texttt{IF-ELSE} conditions,
  \texttt{WHILE} and \texttt{FOR} loops), individual SQL statements (e.g.,
  \texttt{SELECT}, \texttt{UPDATE}, \texttt{DELETE}), variable management,
  exception handling, and others. By capturing the inherent challenges of
  translating procedural logic, SQLProcBench serves as a critical benchmark for
  evaluating the effectiveness of SQL dialect translation approaches in
  practical database migration scenarios.

\end{itemize}

\subsubsection{Methodology} \label{sec:LLMTranslator}

We develop a prototype, \texttt{LLMTranslator}, to evaluate LLM-based
SQL dialect translation and to serve as a baseline for subsequent comparisons.
\texttt{LLMTranslator} adopts the Chain-of-Thought (CoT) strategy \cite{CoT},
executing the translation process in three stages. First, we execute the source
query on both the target and source databases to collect feedback, which serves
as error messages (e.g., execution errors or result inconsistencies). Second, we
use LLMs to identify and summarize dialect-specific features present in the
source query. Finally, LLMs complete the translation by incorporating both the
error feedback and the summarized dialect-specific features.

To further assess the contribution of error messages and dialect-specific
feature summaries to translation performance, we conduct an ablation study on
\texttt{LLMTranslator}. Specifically, we evaluate three variants: one excluding
error messages (w/o \texttt{ErrMsg}), one excluding feature summaries (w/o
\texttt{Summary}), and one omitting both (\texttt{SingleLLM}). To ensure the
generalizability of our findings, we evaluate \texttt{LLMTranslator} using two
representative LLMs: DeepSeek-V3 \cite{deepseek} (denoted as
LLMTranslator$_{DS}$) and GPT-4o \cite{gpt4o} (denoted as
LLMTranslator$_{GPT}$). For each LLM and its variants,
we compute the number of syntax errors and inequivalent queries, allowing a
maximum of three translation iterations per query.

\subsubsection{Metric} \label{sec:metric} The accuracy of translations is
quantified and used as the evaluation metric. We execute the translated query on
the target DBMS and compare its results with those obtained by running the
source query on the source DBMS. If the execution fails or produces inconsistent
outcomes, the translation is deemed unsuccessful. For queries with consistent
results, we further conduct manual verification to confirm semantic equivalence
between the translated query and the source query. The translation is considered
successful only if semantic consistency is verified.

\begin{table}
  \caption{Accuracy of \texttt{LLMTranslator} and Its Variants}
  \begin{tabular}{p{0.245\columnwidth}ccc}
    \toprule
    \textbf{Model} & \textbf{Syntax error} & \textbf{Inequivalent} & \textbf{Accuracy} \\
    \midrule
    \texttt{\textbf{LLMTranslator$_{GPT}$}} & 7 & 12 & 86.71\% \\ 
    \midrule
    w/o \texttt{ErrMsg} & 7 & 18 \color{red}(50.00\% $\uparrow$) & 82.52\% \\
    w/o \texttt{Summary} & 9 \color{red}(28.57\% $\uparrow$) & 17 \color{red}(41.67\% $\uparrow$) & 81.82\% \\
    \texttt{SingleLLM} & 13 \color{red}(85.71\% $\uparrow$) & 21 \color{red}(75.00\% $\uparrow$) & 76.22\% \\
    \midrule
    \texttt{\textbf{LLMTranslator$_{DS}$}} & 9 & 16 & 82.52\% \\ 
    \midrule
    w/o \texttt{ErrMsg} & 10 \color{red}(11.11\% $\uparrow$) & 24 \color{red}(50.00\% $\uparrow$) & 76.22\% \\
    w/o \texttt{Summary} & 12 \color{red}(33.33\% $\uparrow$) & 20 \color{red}(25.00\% $\uparrow$) & 77.62\% \\
    \texttt{SingleLLM} & 14 \color{red}(55.56\% $\uparrow$) & 27 \color{red}(68.75\% $\uparrow$) & 71.33\% \\
    \bottomrule
  \end{tabular}
  \label{tb:empirical}
\end{table}

\begin{figure*}[t]
  \centering
  \includegraphics[width=\textwidth]{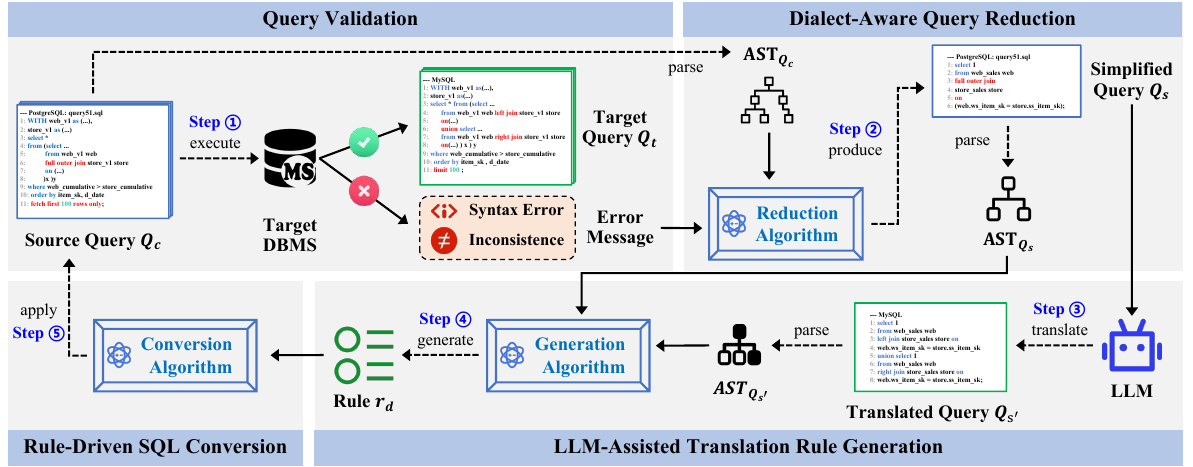}
  \caption{Overview of \approach{}.}
  \label{fig:workflow}
\end{figure*}

\subsection{Translation Accuracy}

As indicated by the empirical results in \tabref{tb:empirical}, even when
using a pure LLM (i.e., \texttt{SingleLLM}) directly, most SQL dialects can be
successfully translated, with GPT-4o and DeepSeek-V3 achieving accuracy rates of
76.22\% and 71.33\%, respectively.

\observation{o1}{A pure LLM alone can correctly translate
the majority of dialects.}

When error messages are available, the LLM can identify the location and the
nature of dialect-specific elements, which enhances the ability to accurately
translate SQL dialects. Similarly, the feature summaries assist the LLM in
achieving a more accurate understanding of the various characteristics within
the source query, thereby reducing erroneous translations. Removing either
the error messages (i.e., w/o \texttt{ErrMsg}) or the feature summaries (i.e.,
w/o \texttt{Summary}) increases the number of syntax errors and inequivalent
queries by approximately 34.09\% and 31.82\%, respectively.

\observation{o2}{Both error messages and dialect-specific feature summaries can enhance the accuracy of SQL dialect translation.}

\subsection{Failure Factors} \label{sec:LLMTranslator-failures} We conduct a
detailed analysis of the erroneous queries produced by
\texttt{LLMTranslator$_{GPT}$} and \texttt{LLMTranslator$_{DS}$}. For these 44
error cases, we first identify the dialect-specific elements. Then, we manually
remove dialect-irrelevant content to create simplified versions and re-translate
these using \texttt{LLMTranslator}.

We find that 38.64\% of the errors are directly resolved after removing
dialect-irrelevant content, as this prevents inconsistencies caused by the LLM
modifying content unrelated to the dialect. For
instance, the LLM alters \texttt{GROUP BY} clauses (e.g., changing \texttt{GROUP
BY} a, b, c to \texttt{GROUP BY} a, b), modifies column names and constants such
as strings within SQL queries, and performs inequivalent optimizations. These
include rewriting expressions like \texttt{Sum(Sum())} to \texttt{Sum()}, as
well as converting implicit joins into explicit \texttt{JOIN} operations, albeit
with incorrect join conditions.

Among the remaining errors, eight were produced by GPT-4o and 19 from
DeepSeek-V3. After applying query reduction, \texttt{LLMTranslator$_{GPT}$}
successfully translated all eight previously error cases. In five of these,
GPT-4o had initially omitted dialects that required handling. These dialects
were correctly processed via query reduction. In the remaining three cases,
GPT-4o was misled by verbose information, resulting in incorrect translations of
the semantically inequivalent \texttt{ROLLUP} function. After reduction,
however, GPT-4o accurately identified the subtle semantic differences and
produced correct translations. For \texttt{LLMTranslator$_{DS}$}, 11 of the 19
translation failures involved incorrect handling of the \texttt{ROLLUP}
function, seven were due to unprocessed dialects, and one involved a
misinterpretation of the \texttt{FULL} \texttt{OUTER} \texttt{JOIN} clause.
Following reduction, all queries except those involving \texttt{ROLLUP} were
successfully translated. These results indicate that \texttt{LLMTranslator} can
effectively handle most SQL dialects in simplified queries, but struggles with
such tasks when dealing with more complex queries. To better understand
\texttt{LLMTranslator}'s limitations, we further analyze the causes of failure
and identify three major types of LLM hallucinations that significantly degrade
translation accuracy.

\begin{itemize}[noitemsep,topsep=0pt,parsep=0pt,partopsep=0pt,leftmargin=15pt]
\item LLMs make improper optimizations in dialect-irrelevant parts.
\item Although LLMs can translate dialects, they often fail to accurately
handle complex contexts.
\item LLMs arbitrarily modify column names, omit certain elements (e.g.,
items in the GROUP BY clause), or alter query objects.
\end{itemize}

\observation{o3}{LLMs exhibit limitations in effectively processing lengthy and complex SQL queries.}

\section{SQL Dialect Translation via Query Reduction}

We propose \approach{}, an automated rule-driven SQL dialect
translation approach based on LLMs. \figref{fig:workflow} shows the overview of
\approach{}. Let $d$ be the dialect that needs to be translated. We start by
validating the source query $Q_{c}$ on the target DBMS (Step
{\color{blue}\textcircled{1}}). If $Q_{c}$ does not produce syntax errors and
yields identical results on its target DBMS, it does not require translation and
is directly returned as the target query $Q_t$. Otherwise, we assume that the
existence of dialect $d$. Then, we perform a dialect-aware query reduction on
$Q_{c}$, in which an adaptive parser converts $Q_{c}$ into its corresponding
abstract syntax tree (AST) $AST_{Q_c}$. Using the error message and $AST_{Q_c}$,
the query reduction algorithm produces a simplified query $Q_{s}$ while
retaining dialect $d$ (Step {\color{blue}\textcircled{2}}). In the next step, we
directly utilize LLMs to translate $Q_{s}$ into $Q_{s^{'}}$ (Step
{\color{blue}\textcircled{3}}). An adaptive parser then converts both the
simplified and translated queries, i.e., $Q_{s}$ and $Q_{s^{'}}$, into ASTs
(i.e., $AST_{Q_s}$ and $AST_{Q_{s^{'}}}$), and a rule generation algorithm
derives the new translation rule $r_d$ of dialect $d$ from these ASTs (Step
{\color{blue}\textcircled{4}}). Finally, we leverage $r_d$ to translate the
dialect $d$ within $Q_{c}$ (Step {\color{blue}\textcircled{5}}) and subsequently
perform query validation on $Q_{c}$. If $Q_{c}$ still contains unresolved
dialect elements or the generated rule fails to translate the dialect $d$ in
$Q_{c}$, we repeat the above process. Otherwise, we return
$Q_{c}$ as the target query $Q_{t}$.

\subsection{Query Validation} \label{sec:validation} We perform query
validation at the following three stages of the process. First, we evaluate the
source query $Q_{c}$ by treating it as a translated query on the target DBMS to
detect its SQL dialect and retrieve error messages. Second, we verify the
LLM-generated translations to ensure that SQL dialects have been accurately
translated. Lastly, we assess rule-driven translations to confirm proper dialect
handling and identify unresolved SQL dialects.

We adopt a dynamic execution strategy to determine whether the translated
query contains unresolved dialect-specific elements and capture associated error
messages. Specifically, we execute both the pre-translation and post-translation
queries on their respective source and target DBMSs and then compare their
outcomes. Before each validation, we reset the database to its initial state,
ensuring consistent evaluation conditions regardless of any intermediate
statements that may have been executed. We conduct validation using the table
data included in the Benchmarks. If the Benchmark does not contain such data,
these data can be generated using existing methods \cite{Houkj, Gray, Bruno}.
The presence of unresolved dialects in the translated query is identified under
the following conditions:
\begin{enumerate}[noitemsep,topsep=0pt,parsep=0pt,partopsep=0pt,leftmargin=15pt]
  \item The target DBMS fails to execute the translated query, and the DBMS
  feedback serves as the error message.
  \item The target DBMS returns different query results from those returned
  by the source query, and the discrepancies between the results serve as the
  error message.
  \item The target DBMS stores different database states after executing the
  query that is intended to modify the database state.
\end{enumerate}

Using this straightforward yet practical method, similar to the approaches
adopted by Mallet \cite{Mallet2024} and VeriEQL \cite{VeriEQL}, we can
effectively verify the syntactic validity and semantic equivalence of the SQL
queries before and after translation.

\subsection{Dialect-Aware Query Reduction} \label{sec:reduction} If
dialects are identified during query validation step, we employ an adaptive
parser to generate the corresponding abstract syntax tree (AST) and perform
subsequent query reduction.

\subsubsection{Adaptive Parsing}

Inspired by SQLess \cite{SQLess}, we implement an adaptive parser using ANTLR
\cite{antlr4} to parse SQL queries into their corresponding ASTs. We
leverage the g4 grammar file \cite{antlr-grammars-v4} of the source DBMS to
generate a basic parser by ANTLR. Then, we utilize the basic parser to parse
the queries into ASTs. If a parsing error
occurs, we will generate a patch for the original g4 grammar
file to complete the parsing rules and resolve the issue. This step is similar to SQLess in achieving an adaptive parser. However,
SQLess does not analyze the semantics of unparseable parts. Instead, it directly adds the unparseable strings into the error patterns of the parsing rule, without creating finer-grained parsing rules for the failed strings. In contrast, we use an LLM, guided by the parsing error message, to analyze the unparseable parts and generate new parsing rules. These rules are then manually validated, and the legitimate ones are integrated into the original g4 grammar file. This approach facilitates the construction of semantically richer patches and the development of a more robust and versatile adaptive parser.

\subsubsection{Query Reduction}

This component focuses on removing as much dialect-irrelevant elements as possible
from the source query while preserving only the dialect-specific elements.
Note that many program reduction methods \cite{HDD, Perses,T-PDD,
LPR, T-Rec, PPR} have been proposed. In our approach, we leverage SQL's AST structure for
syntax-based reduction. Specifically, we adopt a lightweight yet effective method\footnote{In our experiments, our random
reduction strategy takes an average of 16 seconds to reduce a SQL query, achieving a
reduction rate of 91\%.} to randomly delete removable subtrees from the query's AST. Subsequently, we employ an LLM to perform further
query reduction, refining the remaining query structure.

\textbf{Step 1: Syntax-based random reduction.}
We begin by applying random reduction to the
AST parsed from the source query. In each iteration, up to two subtrees are
randomly selected for removal, resulting in a simplified AST and its corresponding simplified query. The resulting simplified query must satisfy two
constraints: (1) It must be successfully executed on the source DBMS; (2) It
must still produce the same error message on the target DBMS. The first
constraint ensures that no new errors are introduced during reduction, while the
second guarantees that the dialect-specific elements are fully preserved. If the
simplified query fails to meet both constraints, the removed subtrees are restored, and alternative subtrees are selected for the next iteration. This process continues until no further subtrees can be removed.

\begin{lstlisting}[caption={Example Prompt for LLM-Based Reduction}, label={listing:simplification-prompt}]
Task description: Migrate the query from {source_db} to {target_db}. Simplify the query below to retain only parts needed to trigger '{original_error}' in {target_db}.
Query: {sql}
CoT instructions: Locate the error, analyze its cause, and remove irrelevant content.
Requirements:
1: Keep the query as short as possible.
2: Using original table/column names.
3: Ensure the simplified query compiles on {source_db}.
4: Preserve the dialect part intact.
Return format:```sql the simplified query here ```
\end{lstlisting}

\textbf{Step 2: LLM-based reduction.} We then apply an LLM to further refine the
query, as random reduction alone may retain dialect-irrelevant elements under
complex constraints (e.g., subqueries and \texttt{CTE}). An LLM is used to
extract and understand the dialect-specific portions, retaining only the
essential parts of the simplified query produced by the random reduction method.
The example prompt is illustrated in \listref{listing:simplification-prompt}. We
avoid using an LLM directly on the source query for reduction, as our empirical
observations indicate that LLMs can introduce various errors when processing
lengthy SQL queries. In this step, the LLM generates five outputs for each
reduction attempt. For each output, we compute a score by executing the
resulting query on the target DBMS, retrieving the new error message, and
calculating the similarity between the original and new error messages. If none
of the outputs exceed a similarity score of 0.85 (an empirically determined
threshold), it suggests that the original dialect has not been fully preserved.
We collect the failed attempts along with their causes of failure and feed this
information back into the next round of the LLM prompting. This process repeats
iteratively, up to a maximum of five iterations. Once an output meets the
criteria, we execute it on the source DBMS. If this execution is error-free, the
output is accepted as final simplified query; otherwise, it is discarded.

\subsection{LLM-Assisted Translation Rule Generation} \label{sec:rule-generation}
This module is divided into two primary components: query translation
and rule generation. It utilizes LLMs to accurately translate the simplified
query and employs a rule generation algorithm to automatically derive
translation rules based on the relationship of both the simplified and
translated queries.

\subsubsection{Query Translation}
After obtaining the simplified query, we utilize \texttt{LLMTranslator}
described in \secref{sec:LLMTranslator} to perform the SQL dialect translation.
The translated query undergoes the validation process outlined in
\secref{sec:validation}. Any translation that fails to execute
or produces results inconsistent with the simplified query is discarded, and
\texttt{LLMTranslator} is re-prompted with the simplified query to generate new
translations.

\begin{figure}[t]
  \centering
  \includegraphics[width=\columnwidth]{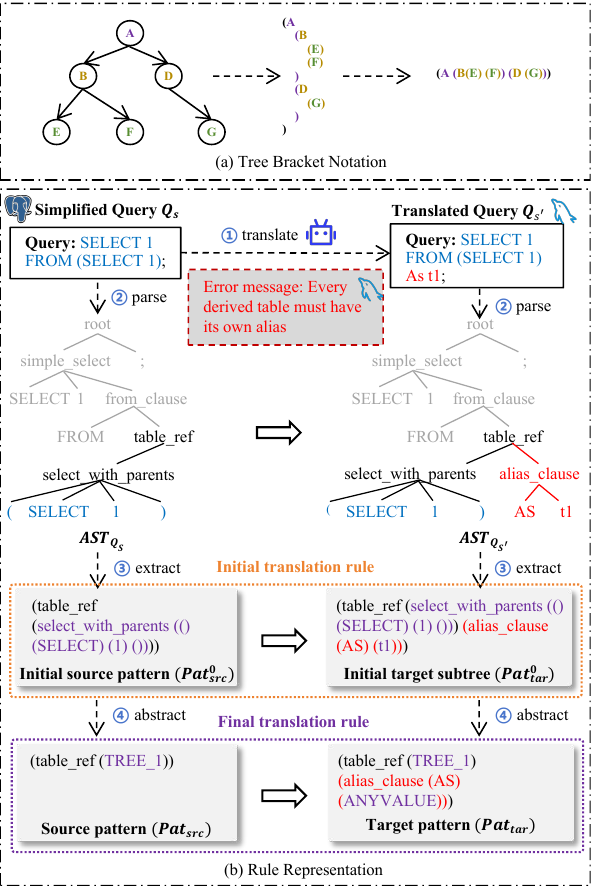}
  \caption{An Illustrative Example of Translation Rule Generation.}
  \label{fig:example-rule-generation}
\end{figure}

\subsubsection{Rule Generation}
This component aims at automatically deriving a translation rule
from the simplified and translated queries.

\textbf{Translation rule definition.}
We define the translation rule as mapping from a source pattern ($Pat_{src}$) to a target
pattern ($Pat_{tar}$), where a
pattern specifies an AST structure representing the syntactic form of the
dialect. Let $AST_{Q_s}$ and $AST_{Q_{s^{'}}}$ be the ASTs of the simplified query and
translated query, respectively. When 
$AST_{Q_s}$ contains a tree structure matching the source pattern, it indicates
a dialect-specific construct requiring transformation. 
SQL dialect translation is achieved by substituting segments of $AST_{Q_s}$ that match the source pattern with the
corresponding target pattern, and this process generates $AST_{Q_{s^{'}}}$.

\figref{fig:example-rule-generation} demonstrates the rule for mandatory alias
assignment to derived tables in MySQL, which requires all derived tables (e.g.,
subqueries) to have explicit aliases. To facilitate representation, we
utilize \emph{bracket notation} \cite{CLRS} to express tree structures. As
illustrated in \figref{fig:example-rule-generation} (a), this notation employs
nested parentheses to represent parent-child relationships between nodes,
thereby uniquely identifying a tree. To extract the rule, we identify the
minimal differing subtrees $T_{src}$ in $AST_{Q_s}$ and $T_{tar}$ in
$AST_{Q_{s^{'}}}$, in which $T_{src}$ and $T_{tar}$ share the same root nodes,
and the remaining parts of the ASTs (i.e., $AST_{Q_s}$.remove($T_{src}$) and
$AST_{Q_{s^{'}}}$.remove($T_{tar}$)) are identical. As shown in
\figref{fig:example-rule-generation} (b), the minimal differing subtrees (i.e.,
$T_{src}$ and $T_{tar}$) correspond to the trees rooted at the
\texttt{table\_ref} node. We perform the translation of dialects by leveraging
the transformations on these two subtrees. Specifically, the subtree $T_{src}$,
extracted from $AST_{Q_s}$, serves as the source pattern of the rule, while the
other subtree $T_{tar}$, extraced from $AST_{Q_{s^{'}}}$, constitutes the target
pattern.

\SetInd{0.12em}{1.0em}
\begin{algorithm}[tb]
  \footnotesize
  \caption{Initial Translation Rule Extraction}
  \label{alg:rule-extraction}
  \KwIn{Simplified query AST $AST_{Q_s}$ and its corresponding translated query AST $AST_{Q_{s^{'}}}$}
  \KwOut{Initial translation rule $\langle Pat^0_{src}, Pat^0_{tar} \rangle$}
  \SetKwFunction{GetMinDiff}{getMinDiffTree}
  \SetKwProg{Fn}{Function}{}{}
  $\langle Pat^0_{src}, Pat^0_{tar} \rangle \gets$ \GetMinDiff{$AST_{Q_s}, AST_{Q_{s^{'}}}$} \\
  \Fn{\GetMinDiff{$T_{src}$, $T_{tar}$}}{
    \uIf{$AST_{Q_s}.remove(T_{src}) \neq
    AST_{Q_{s^{'}}}.remove(T_{tar})$}{
      \Return $NULL$
    }
    \Else{
      $num \gets max(T_{src}.root.childNum, T_{tar}.root.childNum)$ \\
      \For{$i \gets 0$; $i < num$; $i \gets i + 1$}{
        $child_1 \gets T_{src}.root.getChild(i)$ \\
        $child_2 \gets T_{tar}.root.getChild(i)$ \\
        \uIf{$child_1 \neq child_2$}{
          \Return $\langle T_{src}, T_{tar} \rangle$
        }
       \Else{
        $rule \gets$ \GetMinDiff{$child_1.subtree, child_2.subtree$} \\
          \If{$rule \neq NULL$}{
              \Return $rule$
          }
        }
      }
      \Return $\langle T_{src}, T_{tar} \rangle$
    }
  }
\end{algorithm}

\textbf{Rule extraction. }As outlined in \algref{alg:rule-extraction}, \approach{} automatically extracts the initial rule through the following steps:
\begin{itemize}[noitemsep,topsep=0pt,parsep=0pt,partopsep=0pt,leftmargin=15pt]
    \item We utilize the function \texttt{getMinDiffTree()} to extract the
    minimal differing subtrees from $AST_{Q_s}$ and $AST_{Q_{s^{'}}}$, and use
    them as the source and target patterns for the initial rule (Line 1).
    \item In \texttt{getMinDiffTree()}, We first check whether the trees
      derived by removing $T_{src}$ and $T_{tar}$ from the $AST_{Q_s}$ and the
      $AST_{Q_{s^{'}}}$, respectively, are identical (Line 3). If they differ,
      the function directly returns \texttt{NULL}, indicating that the current
      subtrees $T_{src}$ and $T_{tar}$ do not meet the requirements (Line 4).
    \item If the remaining parts of the trees are identical, it indicates
    that the current subtrees $T_{src}$ and $T_{tar}$ meet the requirements. To
    identify any smaller differing subtrees within them, we verify whether all
    child nodes of the root nodes in $T_{src}$ and $T_{tar}$ are identical. If
    any pair of child nodes mismatches (Line 6--10), $T_{src}$ and $T_{tar}$
    are the minimal differing subtrees, and we return them directly
    ({Line 11}).
    \item Otherwise, we recursively call the function on their corresponding
    child subtrees (Line 12--13). If a non-empty result is returned, it is the
    minimal differing subtree pair, and we return it immediately (Line 14--15).
    If all recursive calls return empty values, $T_{src}$ and $T_{tar}$ are
    identified as the minimal differing subtrees and are directly returned (Line
    16). In \figref{fig:example-rule-generation}, this process successively
    identifies the following pairs of subtrees with roots at: <\texttt{root},
    \texttt{root}>, <\texttt{simple\_select}, \texttt{simple\_select}>,
    <\texttt{from\_clause}, \texttt{from\_clause}>, and <\texttt{table\_ref},
    \texttt{table\_ref}>.
\end{itemize}

However, the extracted initial rule in \figref{fig:example-rule-generation}
exhibits limited applicability and struggles to be extended to a broader range
of queries containing the same dialect. The inclusion of dialect-irrelevant
contents in the rule adversely affects its matching conditions. Specifically,
``(SELECT) (1)" in the initial source pattern will restrict the rule to only
match cases where the subquery is ``SELECT 1''. Any variation in the content of
the subquery renders the rule inapplicable. However, the content of the subquery
is irrelevant to the dialect-specific operation of assigning an alias to the
subquery. Regardless of the subquery's internal structure, the alias assignment
remains unaffected. This highlights the necessity of abstracting
dialect-irrelevant content within the rule. Additionally, all SQL statements
require identifiers, such as table names and column names, to be unique within
their respective scopes. In this example, the rule consistently assigns the same
alias `t1' to the subquery. If multiple subqueries in an SQL statement require
aliases, this rule would inevitably cause naming conflicts. Therefore, it is
also essential to abstract the names of newly generated identifiers, ensuring
that each is assigned a unique, randomly generated value.

\begin{algorithm}[t]
  \footnotesize
  \caption{Initial Translation Rule Abstraction}
  \label{alg:rule-abstraction}
  \KwIn{Initial translation rule $\langle Pat^0_{src}, Pat^0_{tar} \rangle$}
  \KwOut{Final translation rule $\langle Pat_{src}, Pat_{tar} \rangle$}
  \SetKwProg{Fn}{Function}{}{}
  \SetKwFunction{Abstract}{abstractRule}
    $Pat_{src} \gets Pat^0_{src}$ \\
    $Pat_{tar} \gets Pat^0_{tar}$ \\
   \Abstract{$Pat^0_{tar}.root$} \\
  \Fn{\Abstract{$curNode$}}{
    \If{$isIdentifier(curNode)$}{
        \uIf{$Pat_{src}.contains(curNode)$}{
          $replaceAll(Pat_{src}, Pat_{tar}, curNode, \$N)$
        }
        \Else{
          $replaceAll(Pat_{src}, Pat_{tar}, curNode, ANYVALUE)$
        }
    }
    \ElseIf{$isConst(curNode)$}{
        \uIf{$Pat_{src}.contains(curNode)$}{
            $replaceAll(Pat_{src}, Pat_{tar}, curNode, \$N)$
        }
    }
    \ElseIf{$isNonTerminal(curNode)$}{
        \uIf{$Pat_{src}.contains(curNode.subtree)$}{
            $replaceAll(Pat_{src}, Pat_{tar}, curNode.subtree, TREE\_N)$
        }
        \Else{
            \ForEach{$childNode \in curNode.getChildren()$}{
                $\Abstract(childNode)$
            }
        }
    }
  }
\end{algorithm}

\textbf{Rule abstraction. }As outlined in \algref{alg:rule-abstraction}, \approach{} uses DFS to traverse all subtrees (Line 17--18), checking if the conditions for three types of abstraction are met.
\begin{itemize}[noitemsep,topsep=0pt,parsep=0pt,partopsep=0pt,leftmargin=15pt]
    \item \textit{Node-level abstraction.} We assume that the names of
    identifiers and the values of constants are dialect-irrelevant. To normalize
    the patterns, we replace identical identifier names or constant values in
    both patterns with the symbol \texttt{\$N} (where \texttt{N} is a
    subscript), which represents an arbitrary value of the same type (Line 5--7,
    10--12). For instance, in the dialect translation "FETCH FIRST 100 ROWS ONLY
    $\rightarrow$ LIMIT 100", the integer value 100 is irrelevant to the
    dialect. Whether the value is 100 or 10, it does not affect the translation
    of this dialect. Therefore, we replace it with \texttt{\$1} to represent an
    arbitrary value of the integer, and gain "FETCH FIRST \$1 ROWS ONLY
    $\rightarrow$ LIMIT \$1".
    \item \textit{New identifier abstraction.} If the target pattern
    contains newly introduced identifiers, we replace them with
    \texttt{ANYVALUE}, which denotes a randomly generated unique value (Line
    8--9). For instance, in \figref{fig:example-rule-generation}, the identifier
    \texttt{t1} is replaced with \texttt{ANYVALUE}.
    \item \textit{Tree-level abstraction.} We assume that subtrees that
    remain unchanged during dialect translation are dialect-irrelevant, so we
    replace them with \texttt{TREE\_N}, which represents an arbitrary tree
    rooted at a node of a specific type (Line 13--15). For example, in
    \figref{fig:example-rule-generation}, the subquery ``SELECT 1'' is
    abstracted as \texttt{TREE\_1}, which represents any subtree rooted at
    `\texttt{select\_with\_parents}'.
\end{itemize}

This two-phase process extracts translation rules by identifying the
minimal differing subtrees under the same node between the source and target
ASTs. It further enhances the generality of these rules by abstracting
dialect-irrelevant content within the patterns.

\begin{algorithm}[t]
  \footnotesize
  \caption{Rule-Driven SQL Conversion}
  \label{alg:sql-conversion}

  \KwIn{Source query's AST $AST_{Q_c}$, translation rule $\langle Pat_{src}, Pat_{tar} \rangle$}
  \KwOut{Converted query's AST $AST_{Q_{c^{'}}}$}

  \SetKwFunction{PatternMatch}{patternMatch}
  \SetKwProg{Fn}{Function}{}{}

  $AST_{Q_{c^{'}}}$ $\gets AST_{Q_c}$ \\
  \ForEach{$subTree \in AST_{Q_c}$}{
    \If{\PatternMatch{$subTree$, $Pat_{src}$}}{
      $newSubTree \gets instantiate(subTree, Pat_{src}, Pat_{tar})$ \\
      $replaceSubTree(AST_{Q_{c^{'}}}, subTree, newSubTree)$
    }
  }

  \Fn{\PatternMatch{$subTree, Pat_{src}$}}{
    \textit{/* Abstract nodes: \$N and TREE\_N */} \\
    \uIf{$isAbstractNode(Pat_{src}.root) \wedge Pat_{src}.root.getType() = subTree.root.getType()$}{
      \Return $True$
    }
    \uElseIf{$Pat_{src}.root \neq subTree.root$}{
      \Return $False$
    }
    \Else{
      $num \gets max(Pat_{src}.root.childNum, subTree.root.childNum)$ \\
      \For{$i \gets 0$; $i < num$; $i \gets i + 1$}{
        $child_1 \gets subTree.root.getChild(i)$ \\
        $child_2 \gets Pat_{src}.root.getChild(i)$ \\
        \If{$\neg$ \PatternMatch{$child_1.subtree, child_2.subtree$}}{
          \Return $False$
        }
      }
      \Return $True$
    }
  }
      
\end{algorithm}

\subsection{Rule-Driven SQL Conversion} 
\label{sec:sql-conversion} 
In this step, we utilize the generated rules to perform SQL dialect translation.
As outlined in \algref{alg:sql-conversion}, we search the AST of the source
query for subtrees that match the rule's source pattern. If a matching subtree
is found, it is replaced according to the target pattern. This process involves
the following steps:

\begin{itemize}[noitemsep,topsep=0pt,parsep=0pt,partopsep=0pt,leftmargin=15pt]
\item \textit{Source pattern matching.} We traverse all subtrees in the original
source query AST $AST_{Q_c}$ to identify those matching the source pattern (Line
2--3). During the matching phase (Line 6--19), we use DFS to traverse the input
subtree and the source pattern, checking whether they are identical (Line
10--19). If the root node of source pattern is an abstract node (i.e.,
\texttt{\$N} and \texttt{TREE\_N}), we consider the match successful as long as
the type of the root node in subtree matches the type of the root node in source
pattern, skipping further recursion (Line 8--9).
\item \textit{Instantiation.} We replace abstract symbols in the target pattern
(e.g., \texttt{\$N}, \texttt{TREE\_N}) with values matched from the source
pattern and substitute \texttt{ANYVALUE} with a random unique value. This step yields a
concrete subtree (Line 4).
\item \textit{Replacement.} We replace the subtree matched by the source pattern with the instantiated new subtree, thereby completing the dialect translation (Line 5).
\end{itemize}

By following these steps, we ensure accurate 
identification and conversion of dialect-specific 
structures within SQL queries, facilitating seamless 
translation across different SQL dialects. 

\section{EVALUATION}
To evaluate the effectiveness of \approach{}, we raise the following four
research questions:
\begin{itemize}[noitemsep,topsep=0pt,parsep=0pt,partopsep=0pt,leftmargin=15pt]
  \item \textbf{RQ3 (Accuracy comparison):} How does \approach{} compare against state-of-the-art baselines in terms of accuracy?
  \item \textbf{RQ4 (Query reduction effectiveness):} How effective is query reduction in terms of accuracy and valid rule generation?
  \item \textbf{RQ5 (Efficiency comparison):} How does \approach{}'s efficiency compare to other LLM-based methods?
  \item \textbf{RQ6 (Generated rule analysis):} What translation rules can \approach{} generate?
\end{itemize}
\subsection{Experimental Setup} \label{sec:experiment-setup}

\subsubsection{Benchmarks} Since existing experimental data
for evaluating LLMs on SQL dialect translation often fails to capture the
complexity of real-world queries \cite{Zmigrod}, we adopt TPC-DS \cite{tpc_ds_3_2_0} and SQLProcBench \cite{plsql} (described in \secref{sec:benchmarks}) as our benchmarks.

\subsubsection{Baselines \& Metrics}
We compare \approach{} against the following six baselines. 
Since Mallet \cite{Mallet2024} is not publicly
available, we are unable to include it as one of our baselines.
We utilize accuracy to evaluate the translation
capability of \approach{}, which is consistent with the metric described in
\secref{sec:metric}.
\begin{itemize}[noitemsep,topsep=0pt,parsep=0pt,partopsep=0pt,leftmargin=15pt]
  \item \textbf{SQLGlot:} SQLGlot \cite{SQLGlot} is an open-source SQL parser
  and transpiler that applies manually crafted rules to translate queries across
  dialects.
  \item \textbf{SQLines:} SQLines \cite{SQLinesMigration} automates schema and
  SQL migration between DBMSs via rule-based query translation.
  \item \textbf{jOOQ:} jOOQ \cite{JOOQSQLTranslation} is a Java library for
  type-safe SQL construction, with built-in dialect translation from Java code.
  \item \textbf{\texttt{LLMTranslator}:} \texttt{LLMTranslator} employs a CoT
  strategy for dialect translation using two distinct LLM configurations: one
  with DeepSeek-V3 (\texttt{LLMTranslator$_{DS}$}) and another with GPT-4o
  (\texttt{LLMTranslator$_{GPT}$}). We present the details in
  \secref{sec:LLMTranslator}.
  \item \textbf{CrackSQL:} CrackSQL \cite{CrackSQL} uses LLMs with cross-dialect
  embeddings and a local-to-global strategy to translate queries by syntactic
  segmentation and matching.
\end{itemize}

\subsubsection{Implementation} \label{sec:implementation}
In the following, we provide implementation details and LLM settings to facilitate reproducibility.

\begin{itemize}[noitemsep,topsep=0pt,parsep=0pt,partopsep=0pt,leftmargin=15pt]
  \item \textbf{Adaptive parser building. }Since PostgreSQL
  \cite{postgres-manual} serves as the source DBMS for all experiments, we
  construct our adaptive parser based on PostgreSQLLexer.g4 and
  PostgreSQLParser.g4 grammar files \cite{antlr-grammars-v4}, and build it with
  ANTLR version 4.13.2 \cite{antlr4}.
  \item \textbf{Construction of TPC-DS. }The official TPC-DS benchmark
  \cite{tpc_ds_3_2_0} lacks PostgreSQL templates. DB2 queries \cite{IBM_DB2} are
  the most similar, with 79/99 compiling on PostgreSQL (vs. 12 on MySQL). After
  manual fixes, we create a PostgreSQL-compatible TPC-DS version in which all
  queries execute successfully on PostgreSQL, while only 12 compile and run
  successfully on MySQL. This adjusted dataset forms our TPC-DS benchmark.
  \item \textbf{LLM settings. }All LLMs used in \approach{} are \textbf{GPT-4o}
  \cite{gpt4o}, a highly advanced LLM. For SQL reduction, the LLM is configured
  with a temperature of 0.7 and a token limit of 2048 to ensure comprehensive
  outputs. In \texttt{LLMTranslator}, the LLM used for feature summary is
  configured with a temperature of 0 and a token limit of 2048 to ensure
  deterministic outputs, while the LLM used for translation is configured with a
  temperature of 0.7 and the same token limit. Both \approach{} and
  \texttt{LLMTranslator} are set to a maximum of three
  iterations. To ensure the reliability of the results, we
  conduct repeated experiments on \approach{} to mitigate the randomness
  introduced by LLMs.
\end{itemize}

\begin{table*}[t]
  \centering
  \caption{Translation Accuracy across Different Methods}
  \begin{tabular}{
    >{\centering\arraybackslash}p{1.5cm}
    >{\centering\arraybackslash}p{1.7cm}
    >{\centering\arraybackslash}p{1.7cm}
    >{\centering\arraybackslash}p{1.7cm}
    >{\centering\arraybackslash}p{2.5cm}
    >{\centering\arraybackslash}p{2.5cm}
    >{\centering\arraybackslash}p{1.7cm}
    >{\centering\arraybackslash}p{1.7cm}
  }
    \toprule
    \bfseries Dataset
      & \bfseries jOOQ & \bfseries SQLines & \bfseries SQLGlot 
      & \bfseries \texttt{LLMTranslator$_{DS}$}
      & \bfseries \texttt{LLMTranslator$_{GPT}$}
      & \bfseries CrackSQL
      & \bfseries \approach{} \\ 
    \midrule
    TPC-DS & 73.74\% & 67.68\% & 75.76\% & 82.83\% & 90.91\% & 80.81\%
        & \textbf{97.98\%} \\ 
    \midrule
    SQLProcBench & 13.64\% & 0.00\% & 0.00\% & 81.82\% & 77.27\% &
        4.55\% & \textbf{100.00\%} \\ 
    \bottomrule
  \end{tabular}
  \label{tb:RQ3-result}
\end{table*}

\subsection{Accuracy Comparison} We conduct a comprehensive comparison
of \approach{} against six baselines (i.e., SQLines
\cite{SQLinesMigration}, SQLGlot \cite{SQLGlot}, jOOQ
\cite{JOOQSQLTranslation}, \texttt{LLMTranslator$_{DS}$},
\texttt{LLMTranslator$_{GPT}$} and CrackSQL
\cite{CrackSQL}) by using two real-world benchmarks
TPC-DS \cite{tpc_ds_3_2_0} and SQLProcBench \cite{plsql}. On TPC-DS,
we translate queries from PostgreSQL \cite{postgres-manual} to MySQL
\cite{mysql2023doc}, while on SQLProcBench, we translate stored
procedures from PostgreSQL to Oracle \cite{oracle2023doc}. For the
three SQL dialect translation tools, we rely on their online
interfaces to manually translate queries. For CrackSQL, we utilize
its open-source implementation, and adopt the default configuration and
knowledge base to ensure a fair comparison.

\subsubsection{Experimental Evaluation}
\figref{tb:RQ3-result} shows the accuracy results of \approach{} and selected baselines in SQL dialect translation. Our experiments reveal the following three-fold key findings.

\sloppy{}\textbf{\approach{} demonstrates superior performance compared to all
  the baselines on the two benchmarks.} Overall, \approach{} achieves an
  accuracy of 97.98\% on TPC-DS and 100.00\% on SQLProcBench. Notably, it
  outperforms the best baseline \texttt{LLMTranslator$_{GPT}$} on TPC-DS by
  7.78\%, and outperforms the best baseline \texttt{LLMTranslator$_{DS}$} on
  SQLProcBench by 22.22\%. These improvements clearly underscore
  the effectiveness of \approach{} in SQL dialect translation task.

\textbf{LLM-based methods achieve better results than traditional automated
  tools overall.} Directly using LLMs for SQL dialect translation (i.e.,
  \texttt{LLMTranslator$_{DS}$} and
  \texttt{LLMTranslator$_{GPT}$}) yields average accuracies of
  86.87\% on TPC-DS and 79.55\% on SQLProcBench,
  demonstrating the promise of LLM-based methods. Besides, CrackSQL achieves
  an accuracy of 80.81\% on TPC-DS but performs poorly on SQLProcBench, with an
  accuracy of only 4.55\%. This poor performance is attributed to its lack of
  knowledge about stored procedures. In contrast, the three
  automated SQL dialect translation tools achieve around 70\% accuracy on TPC-DS
  and struggle with the more complex SQLProcBench. Traditional tools depend on
  manually-crafted rules, which are often incomplete and error-prone. These
  limitations highlight the need for more advanced and adaptable translation
  mechanisms to ensure effective and accurate SQL dialect translation.

\textbf{Advanced SQL features (e.g., stored procedures) significantly influences
  the accuracy of traditional SQL dialect translation tools.} Notably, on the SQLProcBench benchmark, existing tools achieve near-zero accuracy due to the absence of necessary translation rules for processing stored procedures.

\begin{figure}[t]
  \centering
  \includegraphics[width=\columnwidth]{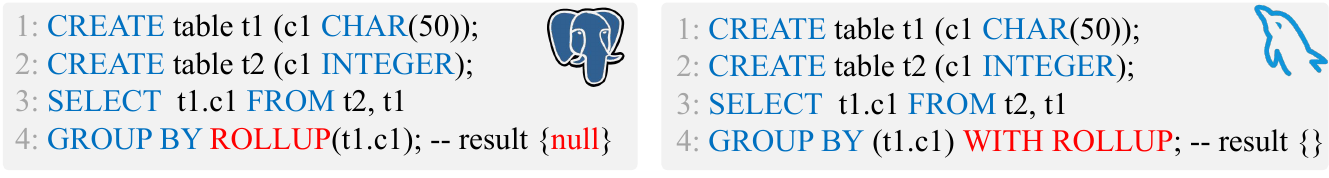}
  \caption{Differences in \texttt{ROLLUP} Semantics between MySQL and PostgreSQL.}
  \label{fig:rollup}
\end{figure}

\subsubsection{Bad Case Breakdown} \label{sec:bad-case}
We further conduct an in-depth investigation into the common error categories observed in the incorrect SQL dialect translations generated by \approach{} and the selected baselines. Specifically, we manually analyze the root causes of each incorrect translation case and categorize them accordingly. Our analysis reveals the following findings based on statistical results.

\textbf{Error analysis for \approach{}.} Only two queries from TPC-DS fail due to semantic inconsistencies. For instance, the symbol `\texttt{||}' acts as a logical \texttt{OR} operator in MySQL but is used for string concatenation in PostgreSQL. 
However, due to the empty result set, \approach{} does not automatically detect this semantic inconsistency and fails to translate them.
Aside from these exceptions, \approach{} successfully translates all other queries.

\textbf{Error analysis for traditional SQL dialect translation
tools.} Our analysis categorizes the failures of traditional tools on the TPC-DS
benchmark into three types: \textbf{parsing errors}, \textbf{incorrect
translation rules}, and \textbf{missing translation rules}. For instance, jOOQ
experiences parsing errors with complex queries (e.g., query 23 exceeding 100
lines), highlighting its limitations with complicated SQL features. Regarding incorrect
translation rules, SQLGlot introduces errors when handling `\texttt{ORDER}
\texttt{BY}' clauses, while jOOQ encounters issues translating the
`\texttt{INTERVAL}' function. In terms of the missing translation rules, SQLGlot
lacks support for assigning aliases to derived tables, jOOQ is unable to process
`\texttt{FULL} \texttt{OUTER} \texttt{JOIN}', and SQLines is deficient in
handling the `\texttt{INTERVAL}' function, `\texttt{NUMERIC}' type conversion,
and `\texttt{FULL} \texttt{OUTER} \texttt{JOIN}' operation. Notably, SQLines occasionally fails to translate the `\texttt{FETCH} \texttt{FIRST} \texttt{n} \texttt{ROWS}' clause, revealing a lack of robustness in its rule for this feature. As demonstrated in \figref{fig:rollup}, the `\texttt{ROLLUP}'
operation exhibits slight semantic differences between MySQL and PostgreSQL, and
none of the three traditional tools process `\texttt{ROLLUP}' correctly,
indicating a broader challenge in handling advanced SQL features. It is
noteworthy that \approach{} effectively addresses these issues with its LLM-assisted translation rule generation module. By leveraging the
extensive domain knowledge of LLM, \approach{} automatically generates
translation rules that overcome the limitations of traditional tools in rule
adequacy. 

\textbf{Error analysis for LLM-based methods.} 
As analyzed in \secref{sec:LLMTranslator-failures}, the errors in
\texttt{LLMTranslator} primarily arise from LLM hallucinations, including
incorrect modifications of dialect-irrelevant content and improper handling of
dialect-specific constructs. \approach{} addresses these challenges via
dialect-aware query reduction. CrackSQL also exhibits significant limitations,
largely due to its heavy reliance on a knowledge base and poor handling of
stored procedures. Its performance on SQLProcBench is severely impacted by
missing stored procedure information, with 68.18\% of errors caused by incorrect
``Cannot translate!'' judgments. On TPC-DS, CrackSQL encounters 19 issues: 1
judgment error, 2 crashes (``list index out of range'' and ``max\_tokens is too
large''), 2 syntax errors, and 14 semantic inconsistencies—all stemming from LLM
hallucinations. For example, CrackSQL incorrectly assumed MySQL does not support
\texttt{INTERSECT} (contradicting the official documentation
\cite{mysql-intersectdoc}) and replaced it with \texttt{UNION}. By removing
dialect-irrelevant elements (e.g., \texttt{INTERSECT}), \approach{} prevents
LLMs from introducing such errors, while query reduction further improves
accuracy on dialect-specific constructs.

\evaluation{c1}{\textbf{Answer to RQ3}: \approach{} significantly outperforms
the six baselines in terms of accuracy, underscoring its effectiveness in
translating SQL dialects.}

\subsection{Query Reduction Effectiveness} \label{sec:RQ4} To assess the impact
of Query Reduction (\texttt{QR}) on \approach{}, we conduct an ablation study on
TPC-DS and SQLProcBench. The study evaluates how removing the \texttt{QR} module
affects \approach{}'s translation accuracy (Acc) and the number of valid rules
(NVR) generated.

\begin{table}[t]
  \centering
  \caption{Ablation Study Results}
  \begin{tabular}{>{\centering\arraybackslash}p{1.6cm}cccc}
    \toprule
    \bfseries Model 
      & \multicolumn{2}{c}{\bfseries TPC-DS}
      & \multicolumn{2}{c}{\bfseries SQLProcBench} \\
    \cmidrule(lr){2-3} \cmidrule(lr){4-5}
     & \bfseries Acc & \bfseries NVR & \bfseries Acc & \bfseries NVR \\
    \midrule
    \approach{} 
      & 97.98\% & 13 
      & 100.00\% & 37 \\
    \midrule
    w/o \texttt{QR} 
      & 81.82\% & 3 & 72.73\% & 20 \\
      & \textcolor{darkgreen}{(16.49\% $\downarrow$)} 
      & \textcolor{darkgreen}{(76.92\% $\downarrow$)} 
      & \textcolor{darkgreen}{(27.27\% $\downarrow$)} 
      & \textcolor{darkgreen}{(45.95\% $\downarrow$)} \\
    \bottomrule
  \end{tabular}
  \label{tb:RQ2-result}
\end{table}

\subsubsection{Results Analysis}
As shown in \tabref{tb:RQ2-result}, removing the \texttt{QR} module (i.e., w/o
\texttt{QR}) significantly impacts the overall performance of \approach{}.
Specifically, without reduction, direct processing of source queries reduces
accuracy by 16.49\% (TPC-DS) and 27.27\% (SQLProcBench), and the number of
generated valid rules drops by 76.92\% and 45.95\%, respectively. The necessity
of \texttt{QR} is twofold.

\textbf{Protection against undesirable modifications.} Without the \texttt{QR}
module, the LLM may inadvertently modify dialect-irrelevant elements during the
translation process. Even if such modifications do not alter the final query's
equivalence, they can limit the general applicability of the generated rules. 
To illustrate, consider a scenario where an LLM unnecessarily adds an `\texttt{AS}' between table names and aliases—a common tendency in its outputs.
If a rule $r$ is to translate dialect $d$, this extra `\texttt{AS}' will limit $r$ to queries in $d$ that lack `\texttt{AS}' before aliases.
More critically,
erroneous modifications (e.g., the example of transforming `\texttt{SUM(SUM())}'
into `\texttt{SUM()}' discussed in \secref{sec:LLMTranslator-failures}) can compromise
query equivalence without triggering compiler warnings, making automated detection of such
semantic inconsistencies extremely challenging.

\textbf{Reduction in query length.} The \texttt{QR} module simplifies the source
SQL queries by removing dialect-irrelevant elements. As observed in \textbf{O3},
LLMs struggle to process lengthy and complex queries. By simplifying the queries, the \texttt{QR} module
enhances the accuracy of LLMs in translating SQL dialects.

\evaluation{c2}{\textbf{Answer to RQ4}: SQL reduction significantly improve the performance
of SQL dialect translation in terms of accuracy and valid rule generation.}

\subsection{Efficiency Comparison}

\begin{figure}[t]
  \centering
  \includegraphics[width=\columnwidth]{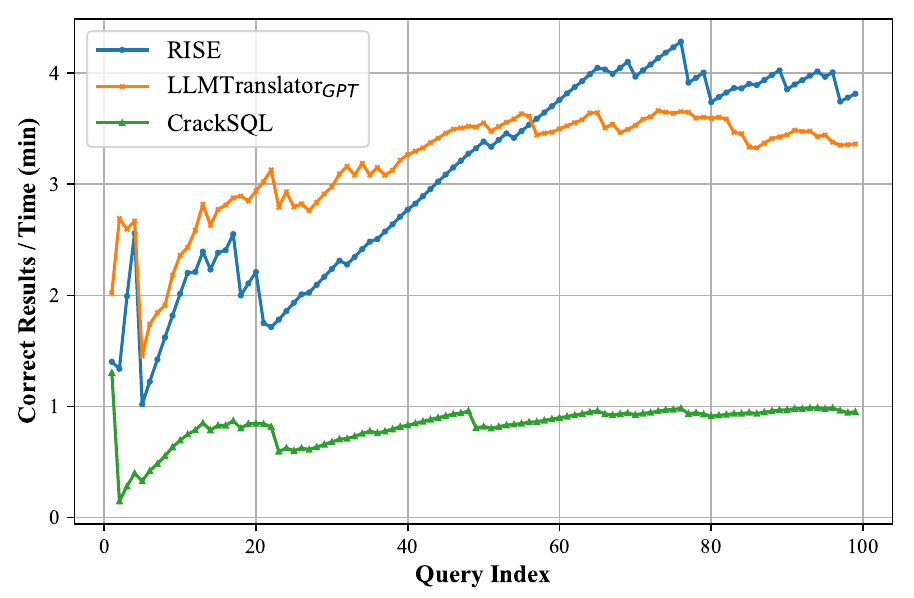}
  \caption{Efficiency Comparison of Correct Translations per Unit Time Between \approach{} and LLM-based methods.}
  \label{fig:RQ3}
\end{figure}

\approach{} has been shown to achieve higher accuracy than other methods. In
light of this, we further investigate the translation efficiency of \approach{}
in comparison with other LLM-based methods (i.e., \texttt{LLMTranslator$_{GPT}$}
and CrackSQL \cite{CrackSQL}) on TPC-DS \cite{tpc_ds_3_2_0}. To ensure fairness,
all methods use the same LLM, GPT-4o \cite{gpt4o}. The benchmarking metric is
defined as follows:

\textit{Correct translations per unit time.} 
Let $n$ denotes the number of correctly translated queries, and $t$ denotes the
total translation time in minutes. The efficiency metric is then measured by
computing the ratio $\frac{n}{t}$. This metric represents the number of correct
translations per minute, providing a quantitative measure to compare the
efficiency of SQL dialect translation.

As shown in \figref{fig:RQ3}, since \approach{} starts with an empty rule set
and has higher complexity, it takes longer to generate valid rules, resulting in
lower efficiency compared to \texttt{LLMTranslator$_{GPT}$}. However, after
accumulating sufficient rules—around query 57—it applies them directly for
translation, achieving a stable rate of 3.78 correct translations per minute,
surpassing \texttt{LLMTranslator$_{GPT}$} at 3.3 translations per minute, which
processes each query independently. The performance gap is expected to grow with
larger datasets. However, due to the high complexity of CrackSQL, its efficiency
has consistently been the lowest. Finally, the total time for \approach{},
\texttt{LLMTranslator$_{GPT}$}, and CrackSQL to complete 99 translations is
25.42 minutes, 26.75 minutes, and 84.02 minutes, respectively.

\evaluation{c3}{\textbf{Answer to RQ5}: Compared to other LLM-based methods, \approach{}
ultimately achieves higher efficiency.}

\subsection{Generated Rule Analysis}

As shown in \tabref{tb:RQ2-result}, \approach{} produces a total of 13 rules for
TPC-DS \cite{tpc_ds_3_2_0} and 37 rules for SQLProcBench \cite{plsql}. All of
the generated rules are confirmed to be valid through manual inspection.

On TPC-DS, \approach{} generates 6 rules to handle dialect-specific constructs: \texttt{FULL OUTER JOIN}, \texttt{INTERVAL}, \texttt{CAST AS NUMERIC/INT}, derived table aliases, and \texttt{FETCH FIRST N ROWS}. On SQLProcBench, it produces 22 rules covering type conversions (\texttt{INT}, \texttt{CHAR}, \texttt{DECIMAL}, \texttt{VARCHAR}, temporal types) and stored procedure constructs (e.g., declarations, cursors, returns, exceptions, IF-ELSE, assignments, LIMIT). Below are two simplified rules generated during evaluation.

\begin{lstlisting}[caption={\texttt{INTERVAL} Function Translation Rule}, label={listing:rule1}]
  source pattern:(...(A_expr_add(TREE_1)(+)(...(Constinterval(INTERVAL))(Sconst(Anysconst('60 days'))))))
  target pattern:(...(Type_function_name(Identifier(DATE_ADD))...(A_expr_add(TREE_1)))(,)(Func_arg_expr(A_expr(INTERVAL)(Interval_value(Iconst(60))(Time_unit(DAY))))))
\end{lstlisting}

\listref{listing:rule1} presents a rule generated by \approach{} in TPC-DS for
translating the \texttt{INTERVAL} function, which existing tools like jOOQ
\cite{JOOQSQLTranslation} and SQLines \cite{SQLinesMigration} fail to handle
correctly. While PostgreSQL requires time values in \texttt{INTERVAL} to be
quoted (e.g., `60 days'), MySQL forbids quotes around such values. However, a simple
string-matching approach risks modifying unrelated string literals resembling
time values. \approach{} overcomes this using the rule-driven SQL conversion
algorithm (\algref{alg:sql-conversion}) to accurately identify \texttt{INTERVAL}
parameters and apply correct conversions.

\begin{lstlisting}[caption={\texttt{Procedure} Declaration Conversion Rule}, label={listing:rule2}]
  source pattern:(Createfunc_opt_item(LANGUAGE)(plpgsql)(AS)(Func_as(Plsql_block((*@\texttt{\$\$}@*))...((*@\texttt{\$\$}@*)))))
  target pattern:(Createfunc_opt_item(IS)(Func_as(Plsql_block ...)))
\end{lstlisting}

\listref{listing:rule2} shows a rule translating PostgreSQL stored procedures to
Oracle by removing ``\texttt{LANGUAGE plpgsql}'', replacing ``\texttt{AS \$\$}'' with
``\texttt{IS}'', and omitting the closing ``\texttt{\$\$}'' delimiter. Specifically,
PostgreSQL uses ``\texttt{LANGUAGE plpgsql}'' to specify the procedural language
and employs ``\texttt{\$\$}'' as the procedure body delimiter, whereas Oracle omits
the language specification and uses ``\texttt{IS}'' to begin the procedure body
without closing delimiters. For example, a PostgreSQL procedure declared as
``\texttt{CREATE OR REPLACE PROCEDURE ... LANGUAGE plpgsql AS \$\$ ... \$\$}'' becomes
``\texttt{CREATE OR REPLACE PROCEDURE ... IS ...}'' in Oracle, ensuring
syntactic compatibility while preserving the procedure's logic.

\evaluation{c4}{\textbf{Answer to RQ6}: On both TPC-DS and SQLProcBench,
\approach{} generates numerous and diverse effective rules.}

\section{DISCUSSION}
In this section, we discuss the threats to validity and the limitations of our approach.
\subsection{Threats to Validity}
The primary threat to the validity of
\approach{} lies in the diversity of the SQL queries used for evaluation and its
generalizability to other DBMSs. In this paper, we collect a total of 143 SQL
queries from TPC-DS \cite{tpc_ds_3_2_0} and SQLProcBench \cite{plsql}, ensuring
a reasonable degree of diversity and quality in our evaluation.
For comparison,
we select three traditional SQL dialect translation tools
\cite{SQLinesMigration, JOOQSQLTranslation, SQLGlot} as well as three LLM-based approaches: CrackSQL \cite{CrackSQL} and our experimental prototype, \texttt{LLMTranslator$_{DS}$} as well as \texttt{LLMTranslator$_{GPT}$}. In future work, we aim to expand the benchmarks for a more comprehensive evaluation of \approach{}'s generalizability across diverse DBMSs and emerging SQL dialects.

\subsection{Limitations}
\textbf{Existence of semantic inconsistency.} Currently, 
\approach{} cannot fully resolve semantic inconsistencies, as relying solely on the execution results of DBMSs cannot ensure complete accuracy. 
For example, as discussed in \ref{sec:bad-case}, \approach{} misidentified the `$||$' operator, which denotes string concatenation in PostgreSQL and logical \texttt{OR} in MySQL. This error introduced a semantic inconsistency between the source and translated queries, leading to translation failure. Moreover, since the data failed to meet the query predicate conditions, both DBMSs returned empty results, causing \approach{} to mistakenly conclude that the queries were semantically equivalent.
It is challenging to automatically validate the semantic
equivalence of the source queries and their corresponding translated queries across DBMSs.
Resolving semantic inconsistency in the SQL dialect translation task remains as
our future work.

\textbf{Inability to handle dialects with ambiguous semantics.} Certain SQL
dialects exhibit semantic ambiguity, as their behavior depends on schema
information from the database and cannot be determined solely from the SQL query
itself. For example, the division operator (`/') in PostgreSQL has two possible
semantics: integer division and floating-point division. The specific behavior
depends on the operands: if the operands include floating-point numbers, it
represents floating-point division; otherwise, it represents integer division.
Since the translation rules currently defined do not account for schema
information, \approach{} is unable to robustly translate queries such as SELECT
t1.c1 / t2.c2, resulting in potentially unreliable translation rules. In the
future, we plan to extend the rules to incorporate schema
information.

\section{RELATED WORK}

\textbf{SQL dialect translation based on manually-crafted rules.} Most current SQL
dialect translation tools
\cite{SQLinesMigration,JOOQSQLTranslation,SQLGlot,Datometry2023}
are rule-driven, leveraging the limited and deterministic nature of SQL dialect
grammars. A straightforward strategy to address recurring error cases is to
design new rules, which can then be integrated into transformation tools to
support comprehensive SQL migrations. However, this method has a clear
limitation: the rules must be manually constructed. This process is not only
time-consuming but also demands a deep understanding of the SQL dialects
involved.

\textbf{LLM-based SQL dialect translation.} LLMs have shown remarkable
code-writing capabilities \cite{Ouyang2023}, providing a promising solution for
SQL dialect translation by reducing manual effort. Mallet \cite{Mallet2024}
leverages LLMs to automatically generate translation rules, thereby reducing the
dependence on manual rule construction for unsupported SQL dialects. However,
the generated rules tend to be relatively simplistic, limiting Mallet's ability
to handle complex SQL features such as stored procedures. Additionally, Mallet
cannot perform SQL dialect translation independently; it relies on the
traditional tool SQLGlot \cite{SQLGlot} for execution, restricting its
applicability to only those DBMSs already supported by SQLGlot. To mitigate
such issues, a recent approach called CrackSQL \cite{CrackSQL} introduces a more
structured framework. It segments complex queries into smaller, functionally
coherent units and applies function normalization and irrelevant query
abstraction to simplify translation, thereby improving robustness and reducing
errors. However, the reliability of LLM-generated outputs remains a significant
concern \cite{Ouyang2023, Weisz2021}. For example, although CrackSQL
\cite{CrackSQL} incorporates LLMs with a knowledge base for validation, our
results reveal frequent syntax errors and semantic inconsistencies.
These findings highlight the limitations of relying solely on LLMs for accurate
SQL translation.

\textbf{Program reduction.} Many approaches have been proposed to address
program reduction from different perspectives. HDD \cite{HDD} and T-PDD
\cite{T-PDD} leverage hierarchical structures and probabilistic modeling to
guide the reduction process. LPR \cite{LPR} introduces LLMs to improve
flexibility and automation. Perses \cite{Perses} and T-Rec \cite{T-Rec} utilize
formal and lexical grammar guidance to ensure syntactic correctness during
reduction. In contrast, PPR \cite{PPR} explores a distinct approach through
pairwise program reduction. Among these, PPR is less suitable for SQL query
reduction due to its reliance on pairwise comparisons, which conflict with the
single-query nature of SQL. The other approaches are more applicable and offer
potential to improve SQL reduction rates. While existing database testing methods
\cite{WriteCheck, DDLCheck, IsoRel} primarily rely on simplistic AST random
deletion, incorporating these techniques into SQL reduction represents a
significant research direction.

\section{CONCLUSION}
In this paper, we begin by exploring the capabilities and limitations of LLMs in
SQL dialect translation through an empirical study. Based on our observations,
we propose a novel approach, \approach{}, for efficiently and accurately
translating SQL dialects across different DBMSs. Specifically, we first apply a
dialect-aware query reduction technique to remove dialect-irrelevant elements
from the source query. We then develop an LLM-assisted translation rule
generation algorithm that automatically creates translation rules based on the simplified query,
enabling effective translation of SQL dialects regardless of query
complexity. Comprehensive experiments on two real-world benchmarks demonstrate the
effectiveness of \approach{}.

\balance

\begin{acks}
This work was partially supported by National Natural Science
Foundation of China (62302493), Basic Research Project of ISCAS
(ISCAS-JCZD-202403), Major Project of
ISCAS (ISCAS-ZD-202302), and Youth Innovation Promotion Association at
Chinese Academy of Sciences (Y2022044).
\end{acks}

\bibliographystyle{ACM-Reference-Format}
\bibliography{mybib}

\appendix
\end{document}